\documentclass[apj]{emulateapj}

\begin{document}
\title{The Nature of the UV/optical Emission of the Ultraluminous X-Ray Source in Holmberg II}

\author{Lian Tao\altaffilmark{1,2}, Philip Kaaret\altaffilmark{2}, Hua Feng\altaffilmark{1}, and Fabien Gris\'e\altaffilmark{2}}
\altaffiltext{1}{Department of Engineering Physics and Center for
Astrophysics, Tsinghua University, Beijing 100084, China}
\altaffiltext{2}{Department of Physics and Astronomy, University of
Iowa, Van Allen Hall, Iowa City, IA 52242, USA}

\shorttitle{UV/Optical from Holmberg II X-1}
\shortauthors{Tao et al.}

\begin{abstract}

We report on UV and X-ray spectroscopy and broad-band optical observations of the ultraluminous X-ray source in Holmberg II.  Fitting various stellar spectral models to the combined, non-simultaneous data set, we find that normal metallicity stellar spectra are ruled out by the data, while low metallicity, $Z = 0.1 Z_{\odot}$, late O-star spectra provide marginally acceptable fits, if we allow for the fact that X-ray ionization from the compact object may reduce or eliminate UV absorption/emission lines from the stellar wind. By contrast, an irradiated disk model fits both UV and optical data with $\chi^2/{\rm dof}=175.9/178$, and matches the nebular extinction with a reddening of $E(\bv)=0.05^{+0.05}_{-0.04}$. These results suggest that the UV/optical flux of Holmberg II X-1 may be dominated by X-ray irradiated disk emission.

\end{abstract}


\section{Introduction}

Ultraluminous X-ray sources (ULXs) are off-nuclear compact X-ray sources whose luminosities exceed the Eddington limit ($3 \times 10^{39}$~erg~s$^{-1}$) of a 20 $M_\sun$ black hole. Thus ULXs may be intermediate-mass black holes that bridge the gap between stellar-mass and supermassive black holes \citep{col99, mak00, kaa01} or stellar-mass black holes with beamed emission \citep{kin01} or in super Eddington accretion states \citep{wat01, beg02}. Please see \citet{fen11} for a comprehensive review.

The accretion disk and the companion star could both contribute to the optical flux of ULXs. Determination of which component is dominant should help understand the physical nature of ULXs. If the optical light is dominated by the companion star, then the flux should be steady and the spectrum should match that of known stars.  In this case, the black hole mass may be obtained via optical radial velocity measurements. If the accretion disk emits most of the optical light, then optical variability is expected and the spectra would likely be similar to those of low-mass X-ray binaries. For a few ULXs, multiwavelength spectra have been employed to distinguish stellar versus disk emission \citep{kaa09,ber10a,gri12}.

Holmberg II X-1 has a bolometric luminosity up to $\sim 3 \times 10^{40}$~erg~s$^{-1}$ \citep{gri10}.  It is surrounded by optical and radio nebulae \citep{pak02,mil05} and has a distance of 3.05 Mpc \citep{hoe98}. The morphology of the optical emission lines demonstrates that the optical nebula is X-ray photoionized \citep{kaa04}. The true X-ray luminosity of Holmberg II X-1 inferred from the flux of high-ionization lines in the nebula, He~{\sc ii} and [O~{\sc iv}], is similar to that estimated directly from the X-ray observations assuming little beaming and is truly in the ultraluminous regime \citep{pak02,kaa04,ber10b}.

A point-like optical counterpart to Holmberg II X-1 is detected within the nebula. Its absolute magnitude $M_V$ and $\bv$ color are similar to stars with spectral types between O4V and B3Ib \citep{kaa04}. Moreover, the UV fluxes observed by the XMM-Newton/Optical Monitor (OM) and GALEX are consistent with a B2Ib supergiant \citep{ber10a}, although due to the poor spatial resolution of the OM and GALEX, the spectral energy distribution (SED) obtained by \citet{ber10a} may be highly contaminated by the nebula and nearby stars (see Figure~\ref{fig:img}).  Holmberg II X-1 exhibits only moderate optical variability, thus a stellar origin of optical flux is consistent with the UV and optical results to date.   However, the optical flux of most ULXs appears to be dominated by an X-ray irradiated disk \citep {tao11}. Here, we present new results on the ultraviolet spectrum of Holmberg II X-1 and combine them with previous results on broadband optical photometry and X-ray spectroscopy to attempt to determine the nature of the optical/UV emission. The observations and data analysis are described in \S~\ref{sec:obs}, and the results are discussed in \S~\ref{sec:dis}.

\section{Observations}
\label{sec:obs}

\subsection{XMM X-ray Data}

Using SAS 11.0.0, we extracted a spectrum from XMM European Photon Imaging Camera (EPIC) PN data obtained on April 10, 2002 (ObsID 0112520601). We found no background flares in the 10-15~keV light curve. The good exposure was 4.7~ks after filtering events with $\rm FLAG = 0$ and $\rm PATTERN \leq 4$. Holmberg II X-1 appears near the CCD gap in the observation, thus the spectrum was extracted in a region centered the source with a moderately small radius of 29\arcsec. Background was estimated using a circular region near the source and subtracted. The spectrum was grouped to have at least 100 counts per bin.

\begin{figure*}
\centerline{\includegraphics[width=\textwidth]{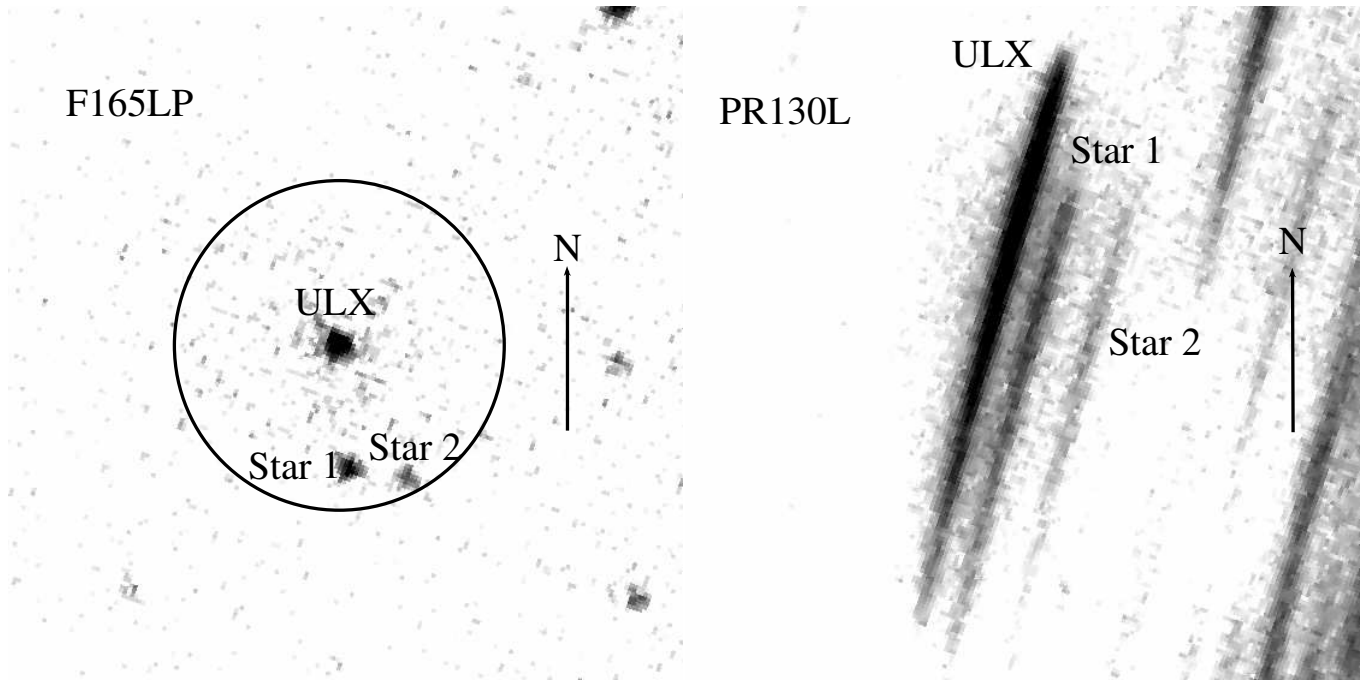}} \caption{HST images around Holmberg II X-1. The left panel is the direct image in the F165LP filter and the right panel is the 2D spectrum observed using the PR130L prism. The solid circle shows the $\sim 1 \arcsec$ spatial resolution of the XMM-Newton Optical Monitor. The arrows have a length of $1 \arcsec$.
\label{fig:img}}
\end{figure*}

\subsection{HST Ultraviolet Data}

We used observations obtained with the Hubble Space Telescope (HST) to avoid nearby stars, and reduce contamination from the nebula. Holmberg II X-1 was observed using the prism grating PR130L of the Solar Blind Channel (SBC) of the Advanced Camera for Surveys (ACS) on November 27, 2006 with an exposure time of 5110~s. For comparison, we also examined an image taken with the F165LP filter (with no prism). The field around the source in the F165LP filter and the PR130L prism are shown in Figure~\ref{fig:img}.

The PR130L prism covers from $\sim 1250$ \AA\ to $\sim 1800$ \AA, and has a nonlinear resolution varying from $\sim 2$ \AA/pixel at 1250 \AA\ to $\sim 20$ \AA/pixel at 1800 \AA. The initial position of the object was taken from the direct image F165LP filter. Then, using the extraction software aXe 2.1 provided by Space Telescope Science Institute (STSI), the spectrum was extracted with a narrow extraction aperture with a width of 8 pixels in order to exclude counts from two adjacent stars, labeled as ``Star 1" and ``Star 2" in Figure~\ref{fig:img}.

In order to check the spectrum extraction, we compared with the source spectrum from the PR130L prism calculated using aXe to the source flux in the F165LP filter at its pivot wavelength calculated using the calibrations in the SYNPHOT package in IRAF.  Unexpectedly, the flux in the F165LP filter was $\sim 30\%$ larger than the flux in the PR130L spectrum from aXe. The same deviation also appears upon checking another source, NGC 1313 X-2. Furthermore, when we extracted a PR130L counts spectrum using aXe then converted it to flux using SYNPHOT, we found that the flux was about 30\% higher than that calculated using aXe directly. Thus we suspect there is a systematic deviation between SYNPHOT and aXe 2.1. The flux conversion of PR130L in aXe is done using sensitivity curve in \citet{lar06}, however, a new ACS SBC throughput curve has been adopted in SYNPHOT \citep{bof08}. Comparing the two curves, there are dramatic deviations at wavelengths below 1270 \AA\ and above 1750 \AA, and the sensitivity at 1250 \AA\ is a factor of $\sim 4$ lower than that at 1300 \AA. Therefore, in this paper, flux conversion was performed using the newest calibration file in SYNPHOT, and we adopted 1300 \AA\ as the short wavelength cut-off, and 1750 \AA\ as a conservative long wavelength cut-off. We noticed that, during the review process of the paper, \citet{fel12} reported the identification of a N~{\sc v}~1240\AA\ line in the same data. This feature is insignificant, more like large fluctuations around a baseline, if we use SYNPHOT to calibrate the flux. The potential large uncertainty of the sensitivity curve prevents reliable detection or characterization of any lines at these short wavelengths.

\begin{figure}
\centerline{\includegraphics[width=\columnwidth]{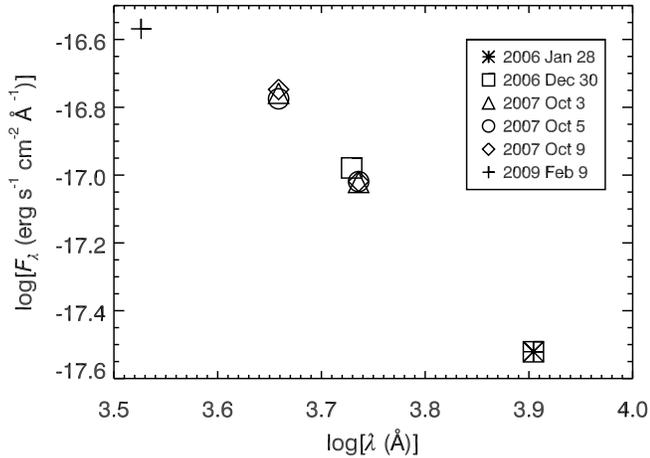}}
\caption{Optical fluxes of Holmberg II X-1 for the available HST observations.
\label{fig:optical}}
\end{figure}

\subsection{HST Optical Data}

There are several HST optical observations of Holmberg II X-1 available. We selected ACS Wide Field Camera (WFC) and Wide-Field Planetary Camera 2 (WFPC2) wideband filter data: ACS F814W (I band) data on January 28, 2006 and December 30, 2006, WFPC2 F336W (U band) data on February 9, 2009; F450W (B band) and F555W data (V band) on October 3, 5 and 9, 2007. Using the APPHOT package of IRAF, we performed aperture photometry with an aperture of $0.2\arcsec$ (4 pixels) on the ACS archival dithered images. For the WFPC2 data, point spread function (psf) photometry was executed on calibrated science (c0f) image using HSTPHOT 1.1\footnote[3]{Available at: http://purcell.as.arizona.edu/hstphot/}. Conversion to fluxes was done using SYNPHOT.

The optical counterpart of Holmberg II X-1 shows moderate variability of no more than 0.07 mag, see Figure~\ref{fig:optical}. Fitting to a constant flux in each band, the $\chi^2$ are 2.8, 3.9 and 0.0 for 3 observations of F450W, 4 observations of F555W and 2 observations of F814W band.  We note that the surrounding optical nebula contributes to the uncertainty in the photometry. \citet{kaa10} found that the BVI fluxes for NGC 6946 ULX-1 increased by about 30\% without nebular background subtraction. Therefore, we compared fluxes with and without nebular subtraction. For F814W, F555W, F450W and F336W band, the flux increases by $\sim$ 17\%, 45\%, 50\% and 27\%, respectively, without nebular subtraction.  Here, we adopt the nebular background subtracted flux with an uncertainty of $\pm20\%$, or $\pm0.2$ mag. This uncertainty is much larger than the time variation of $\leq 0.07$~mag. Thus, using the average flux of these non-simultaneous observations in data fitting is reasonable.  The photometry results in units of erg~cm$^{-2}$~s$^{-1}$~\AA$^{-1}$ are $(2.71\pm0.54)\times10^{-18}$ for F814W, $(7.74\pm1.55)\times10^{-18}$ for F555W, $(1.34\pm0.27)\times10^{-17}$ for F450W and $(2.70\pm0.54)\times10^{-17}$ for F336W.

\begin{figure*}
\centerline{\includegraphics[width=\textwidth]{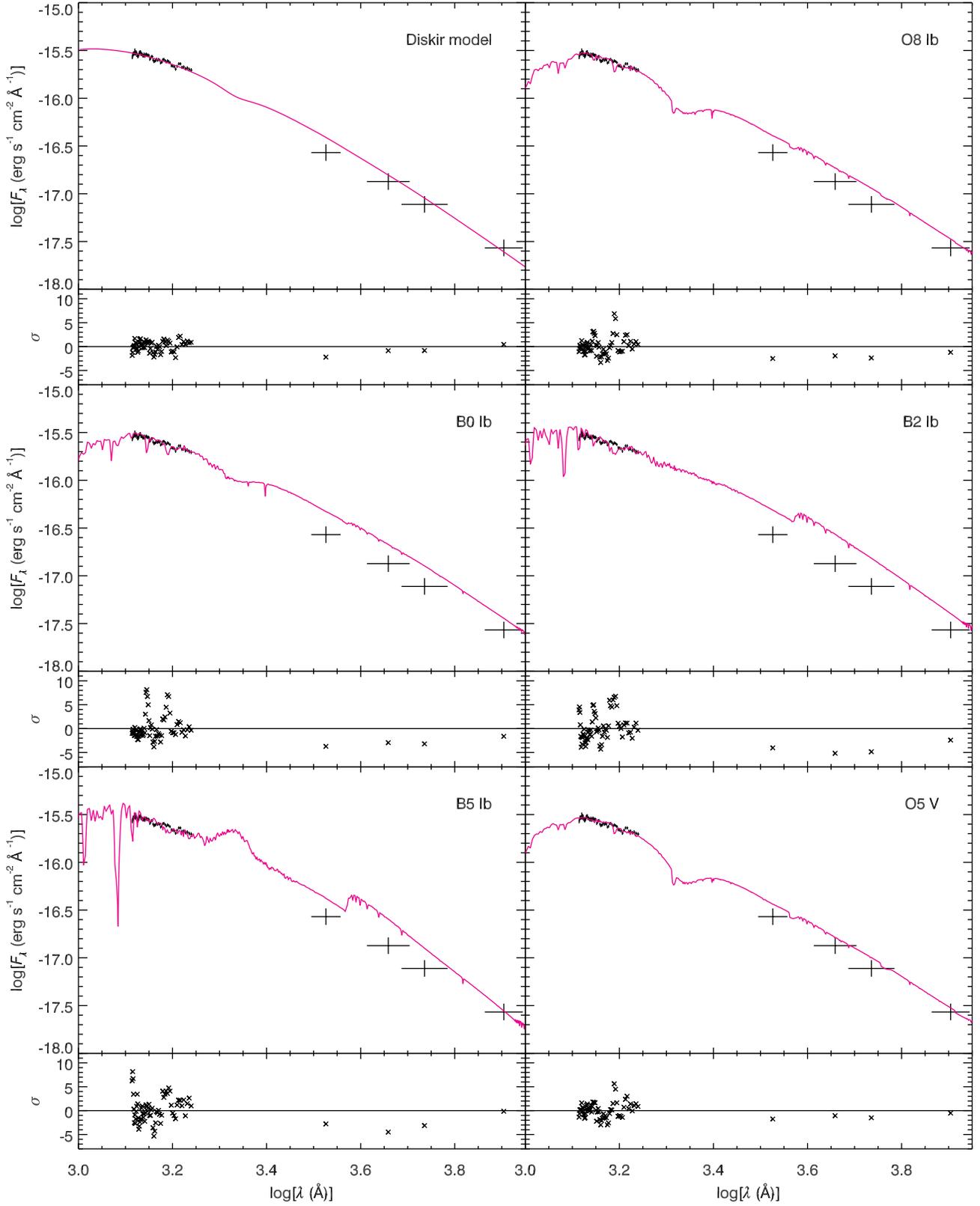}} \caption{UV and optical SEDs with the diskir model and the solar metallicity stellar models for Holmberg II X-1. Upper panels show the absorbed data and models. Lower panels show the residuals in unit of $\sigma$. \label{fig:uvo}}
\end{figure*}

\begin{figure*}
\centerline{\includegraphics[width=\textwidth]{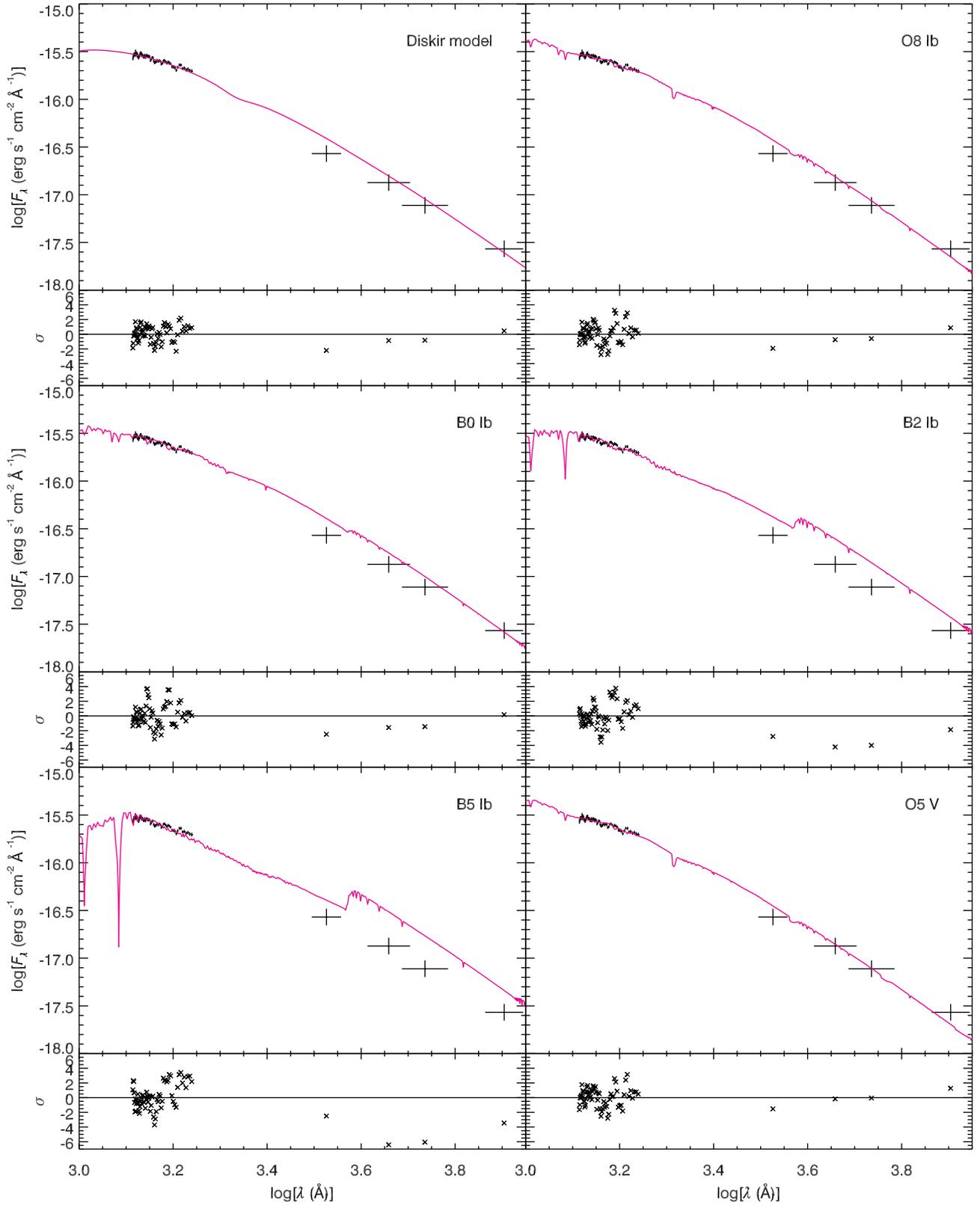}} \caption{UV and optical SEDs with the diskir model and the low-metallicity stellar models.  \label{fig:uvol}}
\end{figure*}

\begin{deluxetable}{llcll}
\tablewidth{\columnwidth} \tablecaption{Spectral Model Fits
\label{tab:stellar}} \tablehead{ \colhead{Model} &
\colhead{$E(B-V)$} & \colhead{Distance} &
\colhead{$\chi^2/{\rm dof}$} & \colhead{$\chi^2/{\rm dof}$} \\
\colhead{} & \colhead{} & \colhead{(Mpc)} & \colhead{} & \colhead{(w/o C {\sc iv})}
}

\startdata
 Diskir  & $0.05^{+0.05}_{-0.04} $  & \nodata & $175.9/178^{\rm a}$ & \nodata \\
\hline\noalign{\smallskip}
\multicolumn{4}{c}{Solar metallicity, $R_V=3.1$}\\
\hline\noalign{\smallskip}
 O8 Ib   & $0.211\pm0.008$   & $2.53^{+0.73}_{-0.60} $ & 232.6/67 & 123.1/56\\
 B0 Ib   & $0.129\pm0.008$   & $2.35^{+0.68}_{-0.55} $ & 476.8/67 & 318.5/56\\
 B2 Ib   & $0.030\pm0.007$   & $2.25^{+0.65}_{-0.53} $ & 583.5/67 & 313.4/56\\
 B5 Ib   & $-0.191\pm0.008$  & $3.41^{+0.99}_{-0.81} $ & 500.9/67 & 372.9/56\\
 O5 V    & $0.230\pm0.008$   & $1.81^{+0.53}_{-0.43} $ & 167.9/67 &  98.8/56\\
\hline\noalign{\smallskip}
\multicolumn{4}{c}{Metallicity $Z=0.1Z_{\sun}$}\\
\hline\noalign{\smallskip}
 O8 I   & $0.103\pm0.003$  &  $3.40^{+0.95}_{-0.77}$  & 116.7/67 &  85.6/56\\
 B0 I   & $0.068\pm0.003$  &  $2.89^{+0.81}_{-0.65}$  & 160.3/67 & 114.7/56\\
 B2 I   & $0.034\pm0.003$  &  $2.34^{+0.65}_{-0.52}$  & 196.5/67 & 121.2/56\\
 B5 I   & $-0.040\pm0.003$ &  $2.34^{+0.65}_{-0.53}$  & 280.8/67 & 223.8/56\\
 O5 V   & $0.111\pm0.003$  &  $2.45^{+0.69}_{-0.56}$  & 105.3/67 &  83.7/56\\
\enddata
\tablecomments{Negative value of $E(\bv)$ means the unabsorbed
stellar model fluxes are redder than the observed optical/UV data. The last column is for fits excluding the C~{\sc iv} region (1500--1600 \AA). $^{\rm a}$~Fitting with both X-ray and UV/optical data.}
\end{deluxetable}

\subsection{Stellar Models}

Inferred from the $M_V$ and $\bv$ quoted in \citet{kaa04}, the optical counterpart of Holmberg II X-1 is consistent with a star with spectral type between O4V and B3Ib. \citet{ber10a} found that the UV fluxes observed by the XMM-Newton/OM and GALEX are too low for an O5V star, but agreed with a B2 Ib supergiant. In their following work, an O5V companion star was excluded because such a star would produce high-ionization line fluxes in the nebula larger than observed \citep{ber10b}.  Relying on the $M_V$ and color of each quasi-simultaneous HST observation discussed above, the most likely stellar type varies from O9Ib to B7I in different observations \citep{tao11}. Here, our UV and optical data are used to test the stellar types of O8Ib, B0Ib, B2Ib, B5Ib and O5V.

Using high resolution X-ray spectroscopy, \citet{win07} found an oxygen abundance for Holmberg II X-1 near the solar value and an iron abundance well above the solar value. In contrast, based on optical spectroscopy of H~{\sc ii} regions, \citet{ric95} found a low oxygen abundance, corresponding to a metallicity of $Z = 0.17 Z_{\sun}$ for the galaxy, where $ Z_{\sun}$ is the solar abundance as estimated by \citet{asp09}. To be conservative, we considered both solar metallicity stellar models and low metallicity, $Z=0.1Z_{\sun}$, stellar models. Referred to the metallicity, effective temperature, and surface gravity \citep{str81,sch82,mar05}, model spectra were selected from the stellar atmospheres models of \citet{cas04}. The model spectra were normalized using the absolute magnitude of the corresponding stellar type from \citet{str81}.

Since different distances have been reported for Holmberg II, e.g.\ 3.05~Mpc in \citet{hoe98} and 3.39~Mpc in \citet{kar02}, distance was adopted as a free parameter, but fits were excluded if the distance obtained was unreasonable. The standard deviation on the absolute magnitude of Galactic O supergiants given a spectral type is 0.45 mag \citep{mar05}. Also, \citet{cro04} reported a scatter of 0.5 mag for OB stars. Thus, a typical uncertainty of 0.5 mag on the normalization was adopted for various spectral types, which was translated into uncertainties of the inferred distances.

We adopted the extinction $E(\bv)$ as another free parameter. For solar metallicity models, the reddening curves of \citet{car89} were used in fitting. $R_V = 3.1$ was assumed in the fitting. However, it could be much larger than that in a particular case. For example, it was found that $R_V = \sim 4 - 6$ for LMC X-1 \citep{bia85,oro09} despite a mean value of 3.1 in the Large Magellanic Clouds \citep{car89}. We thus tested the fits with $R_V$ varying from 3.1 to 6, and the $\chi^2$ increases along with $R_V$ roughly, with $\Delta \chi^2 = 2.7$ (90\% confidence level) at $R_V \approx 3.4$. Therefore, $R_V=3.1$ is preferred and adopted in the fitting. For low metallicity models, we fixed the Galactic extinction to 0.032 \citep{sch98}, and then fit the extinction within Holmberg II adopting the reddening curve of \citet{pre84}.

The spectral energy distributions (SEDs) and the residuals in units of $\sigma$ are shown in Figure~\ref{fig:uvo} and Figure~\ref{fig:uvol}. The extinction, distance and $\chi^2/{\rm dof}$ are reported in Table~\ref{tab:stellar}.

\begin{figure}
\centerline{\includegraphics[width=3.8in]{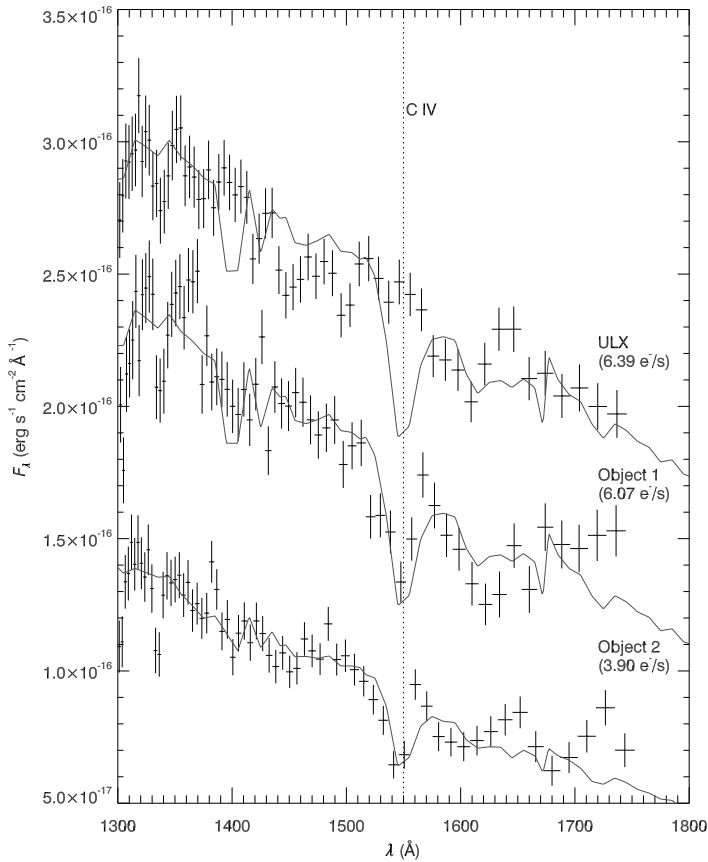}}
\caption{HST prism spectra and best fitting stellar models of Holmberg II X-1 and two comparison objects. The fitting also includes all available photometric data, which are not shown for a clear view of the prism spectra. The spectra of two comparison objects are shifted downward by $0.5\times10^{16}$erg~cm$^{-2}$~s$^{-1}$~\AA$^{-1}$ for clarify. A dashed vertical line indicates the position of the C~{\sc iv} line.
\label{fig:prism}}
\end{figure}

Figure~\ref{fig:prism} shows the prism spectra of Holmberg II X-1 in the wavelength range of 1300 -- 1750 \AA. The spectra of two nearby isolated sources (Object 1: R.A.$=08^{\rm h}19^{\rm m}28^{\rm s}.71$, decl. = +$70^{\circ}42'35.3''$; Object 2: R.A. = $08^{\rm h}19^{\rm m}27^{\rm s}.80$, decl. = +$70^{\circ}42'08.0''$, J2000.0), with a count rate higher than 3 counts s$^{-1}$ in 1300 -- 1750 \AA~ are also plotted for comparison. The $UBV$ colors and magnitudes are consistent with an O8 Ib supergiant for Object 1, and consistent with an O6 V dwarf for Object 2.  They show prominent C~{\sc iv} absorption lines as predicted, while this line is absent in the ULX spectrum, with a 90\% upper limit on the equivalent width of  0.4 \AA\ estimated by fitting the spectrum with an absorption Gaussian with fixed centroid (1550 \AA) and width (14 \AA) plus a power-law continuum with parameters derived outside of the line region (1500--1600 \AA). In Table~\ref{tab:stellar}, we also list the $\chi^2/{\rm dof}$ obtained from fitting the observed spectra with the C~{\sc iv} region removed.

\subsection{Irradiated Disk Model}

To examine the possibility that the UV/optical flux of Holmberg II X-1 originates from an X-ray irradiated disk, we used the irradiated disk model, {\tt diskir} of \citet{gie09}. In the model, X-rays are produced in a standard thin disk and in a Comptonizing corona.  The X-ray bright inner disk illuminates the outer disk and the Compton component illuminates the whole disk outside a given radius.  This illumination increases the temperature of the outer disk and enhances its emission of optical and UV photons.  There are nine parameters in {\tt diskir} model. They are the un-illuminated inner disk temperature $kT_{in}$, the power-law photon index $\Gamma$, the electron temperature $T_e$ of the Comptonizing corona, the ratio of flux between the Compton component and the un-illuminated disk $L_c/L_d$, the fraction of Compton emission thermalized in the inner disk $f_{in}$, the innermost radius illuminated by the Compton tail in terms of the inner disk radius $r_{irr}$, the fraction of bolometric flux thermalized in the outer disk $f_{out}$, the outer disk radius in terms of the inner disk radius $\log(r_{out})$ and the normalization $norm$. We adopted the {\tt wabs} and {\tt redden} models in Xspec to recover the X-ray and UV/optical absorption, respectively. Independent parameters of X-ray absorption versus optical extinction are used because the extinction derived from the X-ray absorption is found to be systematically higher than the extinction inferred from nebular lines for ULXs \citep{tao11}.

\begin{figure*}
\centerline{\includegraphics[width=\textwidth]{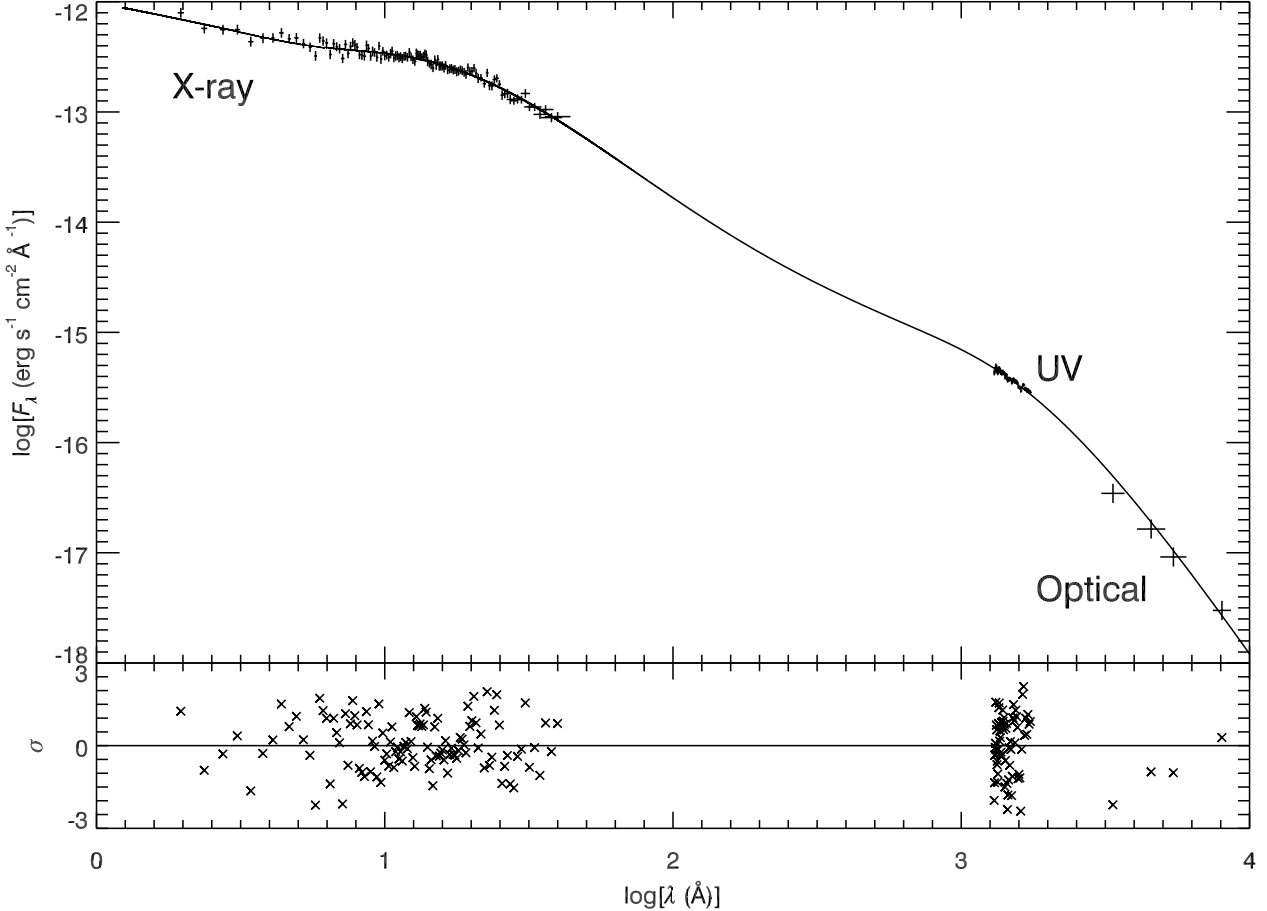}} \caption{The
unabsorbed spectra from X-ray to UV/optical band with the
best-fitting diskir model.
\label{fig:diskir}}
\end{figure*}

$T_e$ and $r_{irr}$ could not be well constrained by the fit, thus we fixed $T_e=100$ keV and $r_{irr}=1.1$ following previous papers \citep{gie09,zur11}. Changing $r_{irr}$ from 1.1 to 1.3, other model parameters change only slightly. Similarly, $f_{in}$ was frozen to 0.1. There are two local minima for $kT_{in}$ in the fit, $\sim0.18$ and $\sim0.10$. The difference in $\chi^2$ between these two minima is about 1.7. We fixed $kT_{in}=0.18$ in the fit, then unfroze it to get its overall uncertainty. A good fit with $\chi^2/{\rm dof}=175.9/178$ was obtained in Xspec. Considering only the UV and optical data, this fit produces a $\chi^2/{\rm dof}$ of 80.3/68. Figure~\ref{fig:diskir} shows the unabsorbed spectra from X-ray to UV/optical band with the best-fitting diskir model. The fitting results are: $E(\bv)=0.05^{+0.05}_{-0.04}$, $N_H=1.08^{+0.14}_{-0.14}\times10^{21}~\rm cm^{-2}$, $kT_{in}=0.09-0.24$ keV, $\Gamma=2.42^{+0.08}_{-0.08}$, $L_c/L_d=1.37^{+0.24}_{-0.19}$, $f_{out}=0.027^{+0.023}_{-0.010}$, $\log(r_{out})=3.51^{+0.06}_{-0.06}$ and $norm=217^{+40}_{-36}$. In order to compare with the stellar models, UV/optical spectra and $\sigma$ with diskir model are plotted in Figure~\ref{fig:uvo} and \ref{fig:uvol}; $E(\bv)$ and $\chi^2/{\rm dof}$ are listed in Table~\ref{tab:stellar}.

In \citet{fen09} and \citet{gla09}, the electron temperatures of Holmberg II X-1 are between 2 and 3 keV. Therefore, we fixed $T_e$ at 2 keV, keeping $r_{irr}=1.1$ and $f_{in}=0.1$, and then fitted the spectrum again. A good fit was obtained, with $\chi^2/{\rm dof}=177.9/177$, and there is only one minimum for $kT_{in}$. The fitting results are: $E(\bv)=0.05^{+0.05}_{-0.04}$, $N_H=0.98^{+0.21}_{-0.18}\times10^{21}~\rm cm^{-2}$, $kT_{in}=0.22^{+0.05}_{-0.04}$ keV, $\Gamma=2.14^{+0.12}_{-0.15}$, $L_c/L_d=0.90^{+0.17}_{-0.13}$, $f_{out}=0.033^{+0.029}_{-0.013}$, $\log(r_{out})=3.68^{+0.18}_{-0.21}$ and $norm=100^{+155}_{-55}$. The main effect of the reduced electron temperature is that the disk normalization decreases by a factor of $\sim2$; other parameters are not affected significantly.

\section{Discussion}
\label{sec:dis}

The results of fitting the UV and optical spectral energy distribution (SED) to stellar models are summarized in Table~\ref{tab:stellar}.  We first consider the continuum fits, excluding the region around the C~{\sc iv} region (1500--1600 \AA), listed in the rightmost column of Table~\ref{tab:stellar}.  The solar metallicity stellar models produce poor fits.  The O5V stellar model produced the best, yet still unacceptable, fit with $\chi^2/{\rm dof} = 98.8/56$, but fitted distance of $1.81^{+0.53}_{-0.43}$~Mpc is unacceptable and \citet{ber10b} find that such a star would over-produce the high-ionization lines observed from the nebula.  The other solar metallicity stellar models produce very poor fits with $\chi^2/{\rm dof} > 2$.  Thus, all the solar metallicity stellar models are excluded.  The low metallicity B-star models are also strongly excluded with $\chi^2/{\rm dof} > 2$.  The low metallicity O5V model produced the best fit, but is, again, excluded by the results of \citet{ber10b}.  The low metallicity O5V model gives a marginally acceptable fit, $\chi^2/{\rm dof} = 85.6/56$ corresponding to a chance probability of occurrence of $7 \times 10^{-3}$.  Thus, we conclude that the stellar models produce at best marginally acceptable fits to the data.

Holmberg II X-1 shows a relatively smooth UV spectrum. In contrast, the stellar models show obvious absorption lines between 1300 \AA\ and 1600 \AA, such as C~{\sc iv} in O8 Ib, C~{\sc iv} and Si~{\sc iv} in B0 Ib,  C~{\sc iv}, Si~{\sc ii},  Si~{\sc iv} and Si~{\sc iii} in B2 Ib,  Si~{\sc iii} in B5 Ib, and C~{\sc iv} in O5 V.  The absence of these lines in the observed spectrum contributes to large $\chi^2/{\rm dof}$ when the full UV spectrum is included in the fits, see Table~\ref{tab:stellar} and Figure~\ref{fig:uvo}.  This effect is strongest for the solar-metallicity models, but still significant for the low-metallicity models.  

However, these stellar UV lines are produced mainly in winds and may be suppressed in the presence of a luminous X-ray source.  X-ray photoionization will either create an ionized Str\"{o}mgen hole around the X-ray source, the Hatchett-McCray effect \citep{hat77}, or completely eliminate the radiatively-driven wind on the X-ray illuminated side of the companion \citep{blo94}. These effects have been observed in high mass X-ray binaries where the UV lines are generated in the wind with a P-Cygni profile \citep{van01,gie08} and should be amplified in ULXs where the ionization degree is orders of magnitudes higher.  Thus, stellar UV lines from ULX binaries should vary with orbital phase unless the binary is nearly face on.  The lines should be absent when we view the X-ray illuminated side of the companion, and appear when we view the part of the stellar wind shadowed from X-rays \citep{gie08}.  Therefore, the absence of C~{\sc iv} lines in a single observation of a ULX cannot exclude a stellar origin of UV emission.  Conversely, if the companions in ULX binaries are high-mass stars, then repeated UV observations should reveal absorption lines and provide information about the binary period.

In contrast to the stellar models, the irradiated disk model fits well to both UV and optical data with $\chi^2/{\rm dof}=175.9/178$ and matches the nebular extinction with a reddening of $E(\bv)=0.05^{+0.05}_{-0.04}$. These results suggest that the UV/optical flux of Holmberg II X-1 may be primarily due to emission from an X-ray irradiated disk.  The best means to definitively determine the physical nature of the emission may be spectral observations around 2000 \AA\ where the deviations between the stellar and disk models are strongest or monitoring observations to search for variability, particularly in the UV.

We note that low mass X-ray binaries, where the UV/optical emission is thought to be dominated by disk irradiation, tend to show significant C~{\sc iv} and other UV line emission due to photoionization \citep[e.g.\ Sco X-1;][]{vrt91}. The UV lines in these sources are thought to be generated in a disk corona where the density is $10^{13}$--$10^{14}$ cm$^{-3}$ \citep{ray93,ko94}.  ULXs with X-ray luminosities near $10^{40}$~erg~s$^{-1}$ may have fully ionized the UV line emitting region, eliminating the UV line emission in a manner somewhat similar to the Hatchett-McCray effect.  Photoionized He~{\sc ii} $\lambda4686$ line emission has been observed in several ULXs, such as Holmberg IX X-1, NGC 1313 X-2, and NGC 5408 X-1, with variable profiles and velocity dispersions up to $\sim 900$ km~s$^{-1}$ suggestive of a disk origin \citep{pak06,cseh11,rob11}.  Other ULXs, like NGC 5204 X-1, show neither He~{\sc ii} nor C~{\sc iv} emission \citep{liu04,rob11}. For Holmberg II X-1, the He~{\sc ii} $\lambda4686$ line is narrow \citep[$\sim 50$ km~s$^{-1}$;][]{leh05} and is consistent with a nebular origin.  He~{\sc ii} emission is produced over a much broader range of ionization states than UV line emission \citep{ray93}, so a search for UV line emission from ULXs showing broad He~{\sc ii} emission might provide information on the profile of X-ray illumination on the disk.

If the UV and optical emission from Holmberg II X-1 are, indeed, due to disk irradiation, then this would suggest that the X-ray emission is due to a standard thin accretion disk.  In this case, the normalization of the thin disk emission is directly related to the size of the accretion disk and, therefore, the black hole mass.  Following \citet{mak00}, we calculate the disk inner radius $R_{in}$ as

\begin{displaymath}
{R_{in} = \xi \kappa^2 {D \over \sqrt{cos(\theta)}}  \sqrt{f_{bol}
\over {2 \sigma T_{in}^4}}}
\end{displaymath}

\noindent where $\xi$ is a correction factor of 0.412 \citep{kub98}, $\kappa \sim 1.7$ \citep{shi95} is the ratio of the color temperature to the effective temperature, $D = 3.05$ Mpc is the distance, $f_{bol} =5 \times \rm 10^{12}~erg~cm^{-2}~s^{-1}$ is the bolometric flux of the multicolor disk model and $\sigma$ is the Stefan- Boltzmann constant.  Using the disk normalization found with the electron temperature fixed to 100~keV, we find $R_{in} = \sim 5 \times 10^8$ cm which corresponds to a mass of $\sim 600M_{\sun}$ for a non-rotating black hole. Using the normalization found for an electron temperature of 2~keV lowers the estimates to $R_{in} = \sim 4 \times 10^{8}$~cm and a black hole mass of $\sim 400 M_{\sun}$.  All mass estimates increase by a factor of 6 if an extreme Kerr black hole is assumed.

The ratio of the outer versus inner disk radius inferred from the {\tt diskir} model is $\sim 3000-7000$.  This is larger than the ratio found in some ULXs, such as NGC 6946 ULX-1, NGC 1313 X-1 and NGC 5408 X-1 \citep{kaa10, yan11, gri12}, but smaller than the ratio $\sim 10^4$ of some Galactic sources \citep{gie09,zur11}.  Using the inner disk radius calculated above, the outer disk radius would then be $\sim 10^{12}$~cm. This is similar to the sizes estimated for the ULXs NGC 1313 X-1 and Holmberg IX X-1 \citep{yan11, tao11}.
The fraction of reprocessed radiation, $f_{out} \sim 0.02-0.06$, is larger than that of Galactic X-ray binaries \citep{gie09}. A larger $f_{out}$ indicates more flux is thermalized in the outer disk.  This could be due to the fact that the disk temperature is lower, and thus produces softer X-rays which are more easily absorbed.  The non-simultaneity of the X-ray versus UV and optical data could also affect the determination of $f_{out}$.

\acknowledgments
We are grateful to the anonymous referee for helpful comments that have improved the paper. LT, PK and FG acknowledge partial support from STScI grant HST-GO-12021. HF acknowledges funding support from the National Natural Science Foundation of China under grant No.\ 10903004 and 10978001, the 973 Program of China under grant 2009CB824800, the Foundation for the Author of National Excellent Doctoral Dissertation of China under grant 200935, the Tsinghua University Initiative Scientific Research Program, and the Program for New Century Excellent Talents in University.

{\it Facility:} \facility{XMM}, \facility{HST}

\end{document}